%% file: ubicomp_v1_main_arxiv.tex
\documentclass[format=acmsmall,preprint]{acmart}

\usepackage{booktabs} % For formal tables

\usepackage{xcolor}	% for highlighting
\usepackage{soul}

\usepackage[ruled]{algorithm2e} % For algorithms

\SetAlFnt{\small}
\SetAlCapFnt{\small}
\SetAlCapNameFnt{\small}
\SetAlCapHSkip{0pt}
\IncMargin{-\parindent}

\setcopyright{none}
\received{July 2017}
\makeatletter
\def\@copyrightspace{\relax}
\makeatother

\begin{document}
% Title portion. Note the short title for running heads 
\title[MobInsight]{MobInsight: A Framework Using Semantic Neighborhood Features for Localized Interpretations of Urban Mobility}  
\author{Souneil Park}
%\orcid{1234-5678-9012-3456} 
\affiliation{%
  \institution{Telefonica Research}
  \streetaddress{Torre Telefónica Diagonal 00 Plaza de Ernest Lluch i Martín, 5}
  \city{Barcelona}
  \postcode{08019}
  \country{Spain}}
\author{Joan Serra}
\affiliation{
  \institution{Telefonica  Research}
  \streetaddress{Torre Telefónica Diagonal 00 Plaza de Ernest Lluch i Martín 5}
  \city{Barcelona}
  \postcode{08019}
  \country{Spain}}
\author{Enrique Frias Martinez}
\affiliation{
  \institution{Telefonica Research}
  \streetaddress{Torre Telefónica Diagonal 00 Plaza de Ernest Lluch i Martín, 5}
  \city{Barcelona}
  \postcode{08019}
  \country{Spain}}
\author{Nuria Oliver}
\affiliation{
  \institution{Telefonica Research}
  \streetaddress{Torre Telefónica Diagonal 00 Plaza de Ernest Lluch i Martín, 5}
  \city{Barcelona}
  \postcode{08019}
  \country{Spain}} 
 
\begin{abstract}
Collective urban mobility embodies the residents' local insights on the city. Mobility practices of the residents are produced from their \textit{spatial choices}, which involve various considerations such as the atmosphere of destinations, distance, past experiences, and preferences. The advances in mobile computing and the rise of geo-social platforms have provided the means for capturing the mobility practices; however, interpreting the residents' insights is challenging due to the scale and complexity of an urban environment, and its unique context. In this paper, we present MobInsight, a framework for making localized interpretations of urban mobility that reflect various aspects of the urbanism. MobInsight extracts a rich set of neighborhood features through \textit{holistic semantic aggregation}, and models the mobility between \textit{all-pairs of neighborhoods}. We evaluate MobInsight with the mobility data of Barcelona and demonstrate diverse localized and semantically-rich interpretations. 
\end{abstract}



% We no longer use \terms command
%\terms{Design, Algorithms, Performance}

\iffalse
\thanks{This work is supported by the National Science Foundation,
  under grant CNS-0435060, grant CCR-0325197 and grant EN-CS-0329609.

  Author's addresses: G. Zhou, Computer Science Department, College of
  William and Mary; Y. Wu {and} J. A. Stankovic, Computer Science
  Department, University of Virginia; T. Yan, Eaton Innovation Center;
  T. He, Computer Science Department, University of Minnesota; C.
  Huang, Google; T. F. Abdelzaher, (Current address) NASA Ames
  Research Center, Moffett Field, California 94035.}
\fi

\maketitle

\renewcommand{\shortauthors}{Park et al.}

\input{ubicomp_v1_introduction}
\input{ubicomp_v1_related_work}
\input{ubicomp_v1_hsa}

\input{ubicomp_v1_nnm}
\input{ubicomp_v1_eval_design}
\input{ubicomp_v1_results_est}

\input{ubicomp_v1_results_int}
\input{ubicomp_v1_conclusion_refs}

\end{document}

%% file: ubicomp_v1_introduction.tex
\section{Introduction}

The mobility practices in urban spaces reflect diverse \textit{spatial choices} stemming from the different lives and experiences of residents. For every spatial choice, consciously or unconsciously, people go through a decision-making process with their insights. They make sense of the context of their travel, consider the spatial and time constraints, project a mental map of relevant areas, and perform a search of a place or a route that best suits their needs. However, the insights involved in this process are not explicitly revealed in the motility itself. The decisions also come more often from intuition than thoughtful reasoning, making it difficult for people to recall and elaborate on them. While such properties hinder the interpretation of the insights, recent advances in mobile computing and geo-crowdsourcing systems are opening new opportunities to approach it at scale, by exposing the movement patterns, detailed information about places, and aggregated preferences and experiences of people.

In this paper, we present MobInsight, a framework for making \textit{localized interpretations} of urban mobility. MobInsight develops an extensive set of local features, and comprehensively explains a fine-grained segmentation of the mobility using the local features. The framework first points out the key features of each neighborhood that affect the mobility of the area. It further expands the view to the relation between all neighborhoods, enabling analyses about how different neighborhood features interact in determining the mobility between them. MobInsight is designed to facilitate site-specific interpretations that are difficult to make with general theories or models of human mobility \cite{erlander1990gravity,simini2012universal,stouffer1940intervening}. Although those general models capture commonly important factors, such as distance and population, they are detached from the unique urban context of the area, which limits the scope of possible interpretations.

Enabling localized interpretations involves challenging problems for research. It requires extensive exploration of numerous possible factors that shape the characteristics of an area, and sophisticated modeling techniques to explain the complex relations between those factors and mobility. MobInsight employs two main approaches to facilitate highly localized mobility interpretations. Firstly, we develop \textit{holistic semantic aggregation} for comprehensive neighborhood feature analysis. It thoroughly identifies the existing places and their function by analyzing diverse online sources, including local guides, geo-crowdsourcing services, also an open directory data from the city government. Having such heterogeneous sources gives a more complete picture of the neighborhoods and mitigates possible selection biases. The method also takes benefit of the semantic annotations left on the places, which capture the meanings given in-situ by actual visitors. Using semantic analysis techniques, the method fuses the structural and linguistic heterogeneity of the annotations across the sources and produces a unified neighborhood profiling scheme. 

Secondly, MobInsight performs \textit{all-pairs inter-neighborhood mobility modeling}, to comprehensively explore the associations between the neighborhood features and the mobility of the people. Having all possible neighborhood pairs in the analysis not only expands the range of potential interpretations, but also mitigates socio-economic, demographic, or regional biases. For this, we take advantage of the real telecommunication logs from the largest operator of the target city, \textit{i.e.}, Barcelona, which enables the construction of a full inter-neighborhood mobility matrix. The logs include 35 million samples of call data records (CDRs) of the residents collected during a month. The data set includes the logs of all mobile phones, not just smartphones or devices with a certain mobile app. The framework employs a multi-layer neural network to learn the complex relationships between the mobility and the features. It also performs \textit{model auditing} for intuitive explanation of feature importances.

The evaluation is composed of two parts. First, through a mobility estimation task, we verify if the neighborhood features contribute to explaining the mobility. Second, we elaborate on the different types of interpretations enabled by MobInsight, and discuss how they are aligned with the descriptions about the urbanism of Barcelona.  

%% file: ubicomp_v1_related_work.tex
% related work
\section{Related Work}
\subsection{Reflections on Urban Mobility}
The rich meanings behind urban mobility have been explored in many studies by reflecting on the psychological, historical, social, and cultural aspects it embodies. As the manifestation of the meanings is inherently implicit, the studies take a larger perspective and consider diverse aspects rather than to simply view mobility as `changes in locations.' The larger perspective allows a deeper interpretation of the mobility with respect to many related themes, such as how people perceive spaces, and build spatial relations and practices. 

Lynch's work \cite{lynch1960image} explores the mental models of urban spaces and investigates the quality of easily recognizable spaces (``legibility'' to use the author's term). While the work implies the importance of the mental representation of the space, recent works have further explored additional factors that affect the perception of the space and mobility such as demographics, technology use \cite{bentley2012drawing}, and emotional pleasantness \cite{quercia2014shortest}. Dourish \cite{dourish2007cultural} emphasizes the diversity in the mobility experience, and challenges the narrow interpretations of mobility that are often found in the mobile computing research of that period. His discussion elaborates on the role of historical and cultural context in how people identify and develop relationship with particular spaces. He also emphasizes the differences in the observations made depending on social groups, even from very similar mobility patterns. The concept `place-identity' \cite{proshansky1983place} of the urbanism literature looks into such particularities in the meanings of places further at the level of individuals, and study how they are related to self-identities. 

We believe De Certeau's concept of ``tactics'' \cite{de1984practice} can be read as an explanation of why such diverse meanings are involved in the mobility. The concept pays attention to the routines of ordinary people who are positioned as `users' of the physical, socio-cultural, and institutional basis, distinguished from those who have the power to shape and control the basis (those who practice ``strategy'' in the author's term). It highlights the creative ways people individualize the basis by altering, adapting, and appropriating it. In the context of urban mobility, he views everyday movements of people as appropriation of the physical space. The space obtains individualized meanings through the spatial tactics of people, which often go beyond rules or expected uses of the space.

We share the view rooted in the above works that rich meanings are involved behind the exhibited mobility, which emphasizes the importance of the mobility studies that are specific to the site and context of the target area. We believe the recent growth of geo-social web services and open data initiatives are creating new opportunities for such studies. Our work aims to provide the tools and techniques for these studies, empowering them to deal with the diversity and complexity of the information involved.

\subsection{Computational Approaches to Mobility Interpretation}
A large body of work exists on data-driven analysis of human mobility due to the pervasive use of mobile devices and the growing adoption of geo-crowdsourcing applications. Focusing on the works that attempt to explain human mobility, we put the relevant works broadly into two classes. A class of works explore common factors behind human mobility. A frequent topic of the works is to develop estimation models for a mobility application, such as commute patterns \cite{lenormand2015comparing}, international trade \cite{koo1991determinants}, and virus spreading \cite{frias2011agent}. As the works develop generalized models, many of them build upon the laws of physics, for example, the gravity model \cite{erlander1990gravity}, which uses the distance between two points and the `mass' of those points (\textit{e.g.}, a property such as population). There are also works that use the models that consider the availability of opportunities, such as the radiation model \cite{simini2012universal} or Stoufer's law of intervening opportunities \cite{stouffer1940intervening}. The availability of opportunity is approximated often with the number of jobs or existing places in an area \cite{noulas2012tale}. 

We conjecture that another direction of work is to look for local, site-specific interpretations of the mobility. However, there are relatively fewer works of this line despite the diversity of cities with their own urban context. Although there are works that use a city-specific data set (\textit{e.g.}, social media check-ins or mobile communication logs of a city) the findings are often made over a common interpretation frame which does not sufficiently capture the local uniqueness of the city; for example, many works \cite{lenormand2015comparing, yuan2012discovering} have studied the functional areas of a city according to the general land use classes, such as residential, business, entertainment, etc. Similarly, the works on spatio-temporal patterns \cite{kling2012city,long2012exploring} often observe diurnal cycles or major activities that can be found similarly in different cities. Our view is that it is important to make more localized interpretations that reflect on the historic, economic, and cultural context of the site, and it is necessary to have tailored techniques or analysis methods for the purpose. 

We believe that Cranshaw et al.'s work \cite{cranshaw2012livehoods} shares our view since it explores the space of interpretations specific to the selected site. The work takes Foursquare check-ins in Pittsburgh and identifies geographical clusters of Foursquare venues based on the check-in patterns. The validation explores the relation between the identified clusters and various factors that shape the dynamics of the city, such as the economic background of areas, demographics, and administrative boundaries, and geography. Our work shows that local insights can be also obtained through a very different analysis. In addition, the task and the data set we develop enables a quantitative assessment for certain aspects of the framework, whereas the cluster analysis of \cite{cranshaw2012livehoods} had to be fully qualitative. 

\subsection{Data-driven Analysis of Urban Spaces}
In a larger context, our work is related to the emerging area of urban informatics. New types of digital data on urban spaces have encouraged many works to study various aspects of an urban area. The aspects explored include the ones typically studied in offline surveys and also those that became newly measurable through the new data (e.g., walkability \cite{quercia2015digital}). For example, Smith et al. \cite{smith2013finger} studied the deprivation status of the areas in London, and analyzed if associations can be found with the usage of public transportation. Call logs data was also used to estimate the liveliness of areas in De Nadai et al.'s work \cite{de2016death}. They explored associations between the liveliness of neighborhoods and a number of properties related to diversity (types of buildings, streets, density, etc.) to quantitatively evaluate the associations suggested in an earlier study of Jane Jacobs \cite{jacobs1961death}. In addition, the photos shared in Flickr were used together with the Foursquare venues to understand various aspects of walkability of the streets in London \cite{quercia2015digital}.

While geo-tagged social media is frequently used in many works, a few recent works take a step back and evaluate the validity of the data sources. Johnson et al. \cite{johnson2016geography} assess the assumption that geo-tagged social media data are made by local people. They observe that such an assumption does not hold for a significant amount of the data and further find socio-demographic biases. Another work of Johnson et al. \cite{johnson2016not} looks into the urban/rural divide in OpenStreetMap and Wikipedia places, and observe similar socio-demographic biases and quality differences. These findings resonate the concerns expressed in an earlier work \cite{dourish2007cultural}, which pointed out the limited view of expected users found in many mobile computing research (often young, affluent and familiar with technology). Though it is difficult to fully address such concerns when using newly emerging social platforms, we acknowledge the possible limitations in our study and try to mitigate them by expanding the collection of data to a broad range of sources of different characteristics.

%% file: ubicomp_v1_hsa.tex
\section{MobInsight Framework}
\begin{figure}
  \includegraphics[scale=0.4]{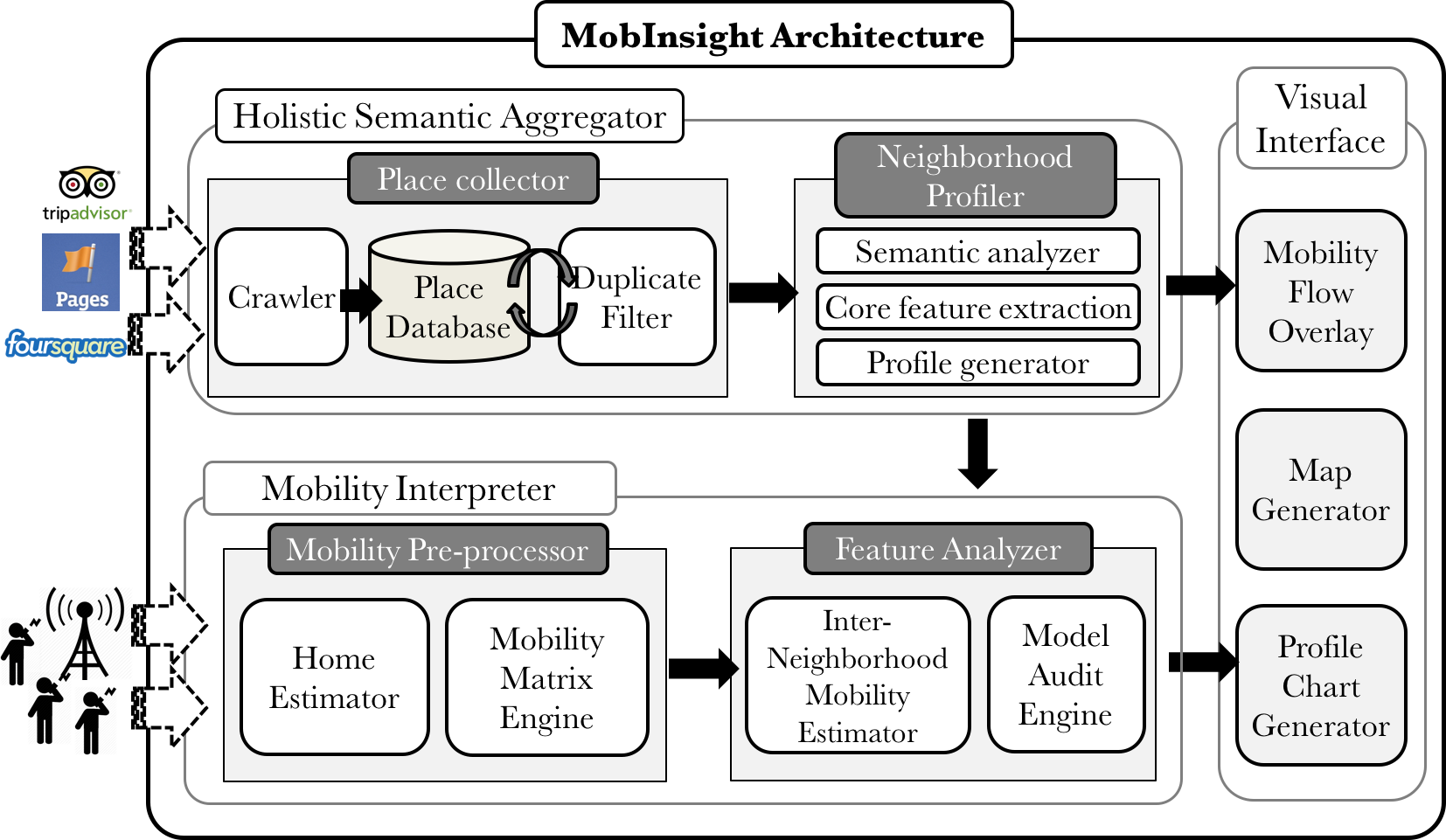}
  \caption{MobInsight Architecture.}
  \label{fig:one}
\end{figure}

Figure~\ref{fig:one} shows the architecture of the MobInsight framework. The framework runs two main data processing flows; first, the one that implements our holistic semantic aggregation approach, collecting places of all neighborhoods and computing their profiles through semantic analysis; second, the flow which uses the neighborhood profiles for mobility modeling and feature analysis. The results of the two data processing flows are merged into a visual interface, which provides an integrated view of the features and mobility, and supports interactive exploration. In this section, we describe the main techniques of the two data processing flows. The visual interface is explained later in the evaluation section together with example mobility interpretations.

\subsection{Holistic Semantic Aggregation}

Although advances in mobile computing and geo-social platforms have produced useful tools for understanding urban spaces, many challenges and issues arise if an analysis has to comprehensively look at the details. While diverse potential data sources are available, they are extremely heterogeneous. They focus on different aspects of a city as they have different goals; for example, Wikipedia and traveling guides cover different places and offer different type of information. In addition, the structure and the language that describe the places vary among the sources. 

Our basic intuition is that the different character of neighborhoods can be captured if the existing places and their function can be comprehensively identified and aggregated. For example, a neighborhood with many handicraft shops and art galleries is likely to have a different character than a neighborhood with department stores. The intuition is connected to a number of urbanism theories that also use the places and their meanings as essential elements for characterizing an area. Place-identity and placemaking studies \cite{proshansky1983place} put the perception and experience of people in the places at the center of understanding an area. These theories not only highlight the importance of the places but also imply the potential value of crowd sourced descriptions. The places are also basic elements in urban morphology \cite{moudon1997urban} and space syntax \cite{hillier2007space} theories, though they put more emphasis on the layout and connections between them as they focus on the urban structure and its development. Empirical qualitative studies \cite{naess2002urban} that observed the influence of urban facilities on mobility more directly also motivates a large-scale data driven analysis. 

To achieve the goal, holistic semantic aggregation covers diverse sources for place identification, fuses the heterogeneous descriptions, and creates a unified aggregation scheme for neighborhood profiling. Due to the extensive coverage of existing places, the method produces a neighborhood profiling scheme that is much detailed than conventional schemes used in urban planning (\textit{e.g.}, residential, business, commercial, entertainment, etc.). As the profiling scheme is built by taking rich semantic metadata of the places and applying linguistic techniques to it, the scheme also allows intuitive interpretations. 

We now describe the two steps of the method in detail: \textit{place identification}, and \textit{semantic aggregation}.

\paragraph{\textbf{Place Identification}} 
We explore various available online resources for data collection and choose 15 different sources, including geo-social services, traveling/local guides, mapping services, and open directory data of the local government. Table~\ref{tbl:one} lists all the sources used for the collection and the number of places collected from each source. If available, the data collection was conducted through the REST API offered by the source. As for the others, we built a scraper customized for each source. The final collection has more than 128,000 places. The collected information includes the name and the location (address and coordinates), source specific meta-data (e.g., number of check-ins, stars), and reviews left on those places. 

Since many sources have tourists as their target audience, a possible bias that we considered in the collection is the tourist bias, especially given the importance and the size of the tourism sector in Barcelona. Another possible limitation is the bias to commercial places, since the sources are mostly connected directly or indirectly to businesses. Regarding the geo-social services, it is possible that they might not include enough routine places (\textit{e.g.}, groceries, work, school) as the users have less motivation to share the locations in such places \cite{lindqvist2011m}. These biases can result in the exclusion of important places that actually have a strong relation to mobility. 

The open directory data of the local government \cite{OpenBCN} was added especially to mitigate the aforementioned biases. The data is produced through a manual survey of all the addresses registered to the city except private homes. As shown in Table~\ref{tbl:one}, a relatively small portion of places of this source overlaps with those of other sources, which suggests that the source is covering different aspects of the city. We also conducted a quick analysis of the category tags and observed that public sector venues and many small businesses (\textit{e.g.}, retail, repair shops, pharmacies) are listed only in the government data. 

\begin{table}
  \includegraphics[scale=0.4]{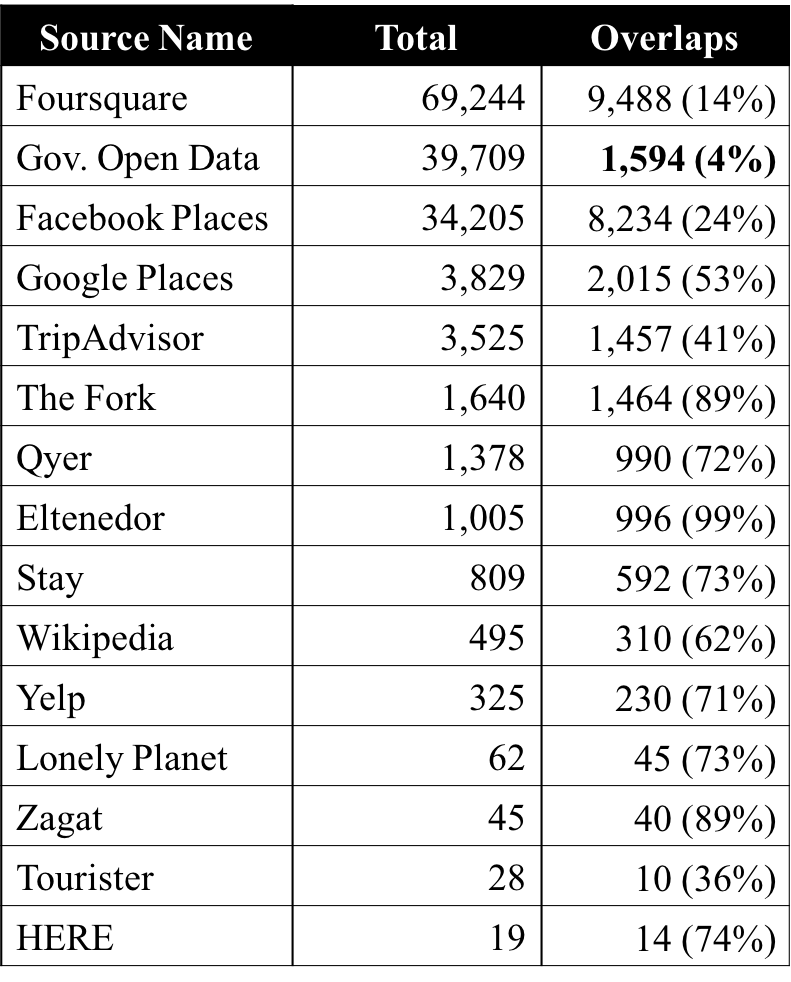}
  \caption{Data sources and the number of collected places.}
  \label{tbl:one}
\end{table}

A challenge that arises from using multiple sources is the existence of duplicates across the sources, especially for famous places. Resolving duplicates is important as they distort the neighborhood profiles. However, identifying duplicates is not a trivial task, since the properties of places that hint the resolution (\textit{e.g.}, the name and coordinates) slightly vary across the sources for the same places. For example, some sources list the same place `Mobile World Center' with a slight variation, \textit{e.g.} `Mobile World Center, Barcelona.' In addition, popular names of an area, such as the metro station name, appear in many different places in the area, which confuses the identification of the name variations. 

In principle, we resolve duplicates by merging the places that have overlapping tokens in their names and are located in close proximity. However, in order to avoid merging different places that simply share a common token in their names, we set a threshold for the candidate places that share the token and do not perform the merging if the number is greater than the threshold. Since an identical place cannot appear more times than the number of data sources, we set the threshold to the number of sources used. The places whose coordinates are located within 50 meters are finally merged. 

\paragraph{\textbf{Semantic aggregation}} 
This step creates a categorization scheme of places specific to the city of interest, and completes the profiling by aggregating the places of each category. The ultimate profile allows computing the difference between neighborhoods by comparing the number of places under each category. The core of this step is creating a categorization scheme that reflects the diversity of the collected places, while keeping the number of categories to a reasonable amount for intuitive interpretation. In order to extract a diverse and inclusive set of categories, we take all the places in the entire city into account, not just those of a specific neighborhood.

The categories are extracted by applying semantic analysis techniques to the meta-data of the places including the tags (crowd-sourced keywords) and the classification taxonomy of the sources. This creates the challenge of dealing with the wide-variety of vocabularies used in the meta-data. The crowd-sourced tags are less structured and the taxonomies are different across the sources. 

We apply a combination of dimensionality reduction and clustering to obtain a set of categories that preserve the diversity and semantic relatedness of the meta-data. First, the meta-data of each place is mapped to a binary word vector by taking all possible n-grams up to the number of tokens, and applying lemmatization to them. Considering the type of data (\textit{i.e.}, set of keywords) we chose latent semantic analysis \cite{deerwester1990indexing} for dimensionality reduction. The number of dimensions is reduced to 100, which explains 77\% of the variance.

K-means clustering \cite{xu2008clustering} is applied over the reduced dimensions to categorize the places. The number K is chosen empirically based on the silhouette score, which measures the consistency of clusters by computing the distance of elements to their own cluster compared to the other clusters \cite{rousseeuw1987silhouettes}. The score increased with K and saturated around 0.7 for K larger than 85, thus we take K=85. 

Instead of taking the 85 clusters as the final categories of places, we tried to further reduce the categories to a reasonable number which allows intuitive interpretation of the neighborhood profiles. We went through the most frequent terms of the clusters manually, and merged the clusters that seem redundant or those that can be combined under a higher level of abstraction (\textit{e.g.}, merging the clusters `financial services' and `advertisement agency' under the abstraction `professional service'). 

\begin{table}
  \includegraphics[scale=0.4]{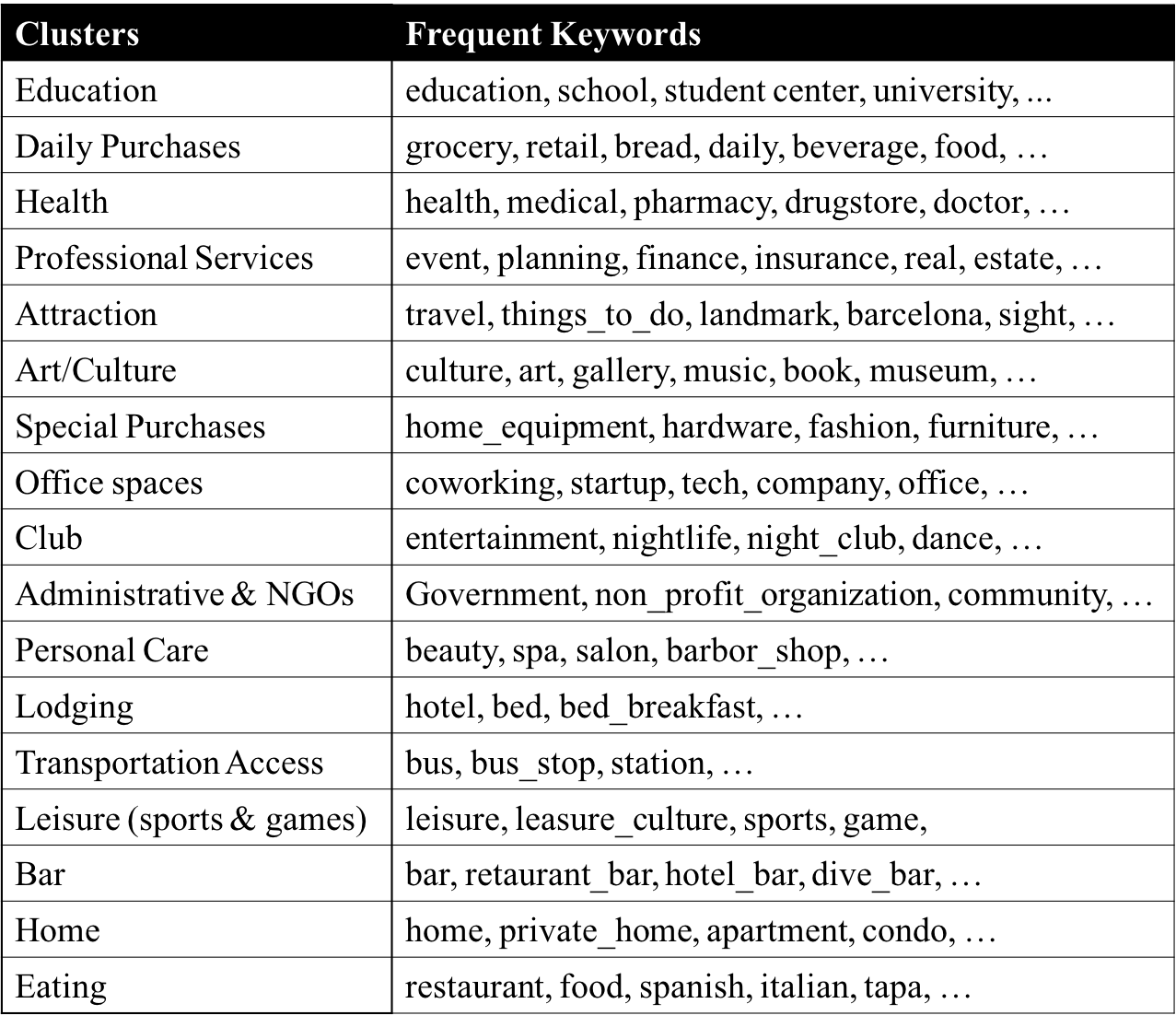}
  \caption{Categories from the clustering result.}
  \label{tbl:two}
\end{table}

We believe there is a possible trade-off between having many fine-grained categories and interpretability, and identifying the optimal granularity is an unclear problem which may depend on many factors such as the goal of an application, local context of the city, etc. As the focus of our current work is on the development of the overall framework and its evaluation, we first use the 17 categories shown in Table~\ref{tbl:two}, which were produced through the above process. We add the total place count as an additional category, thus, 18 features are ultimately used to profile the neighborhoods. 

\begin{figure}
  \includegraphics[scale=0.25]{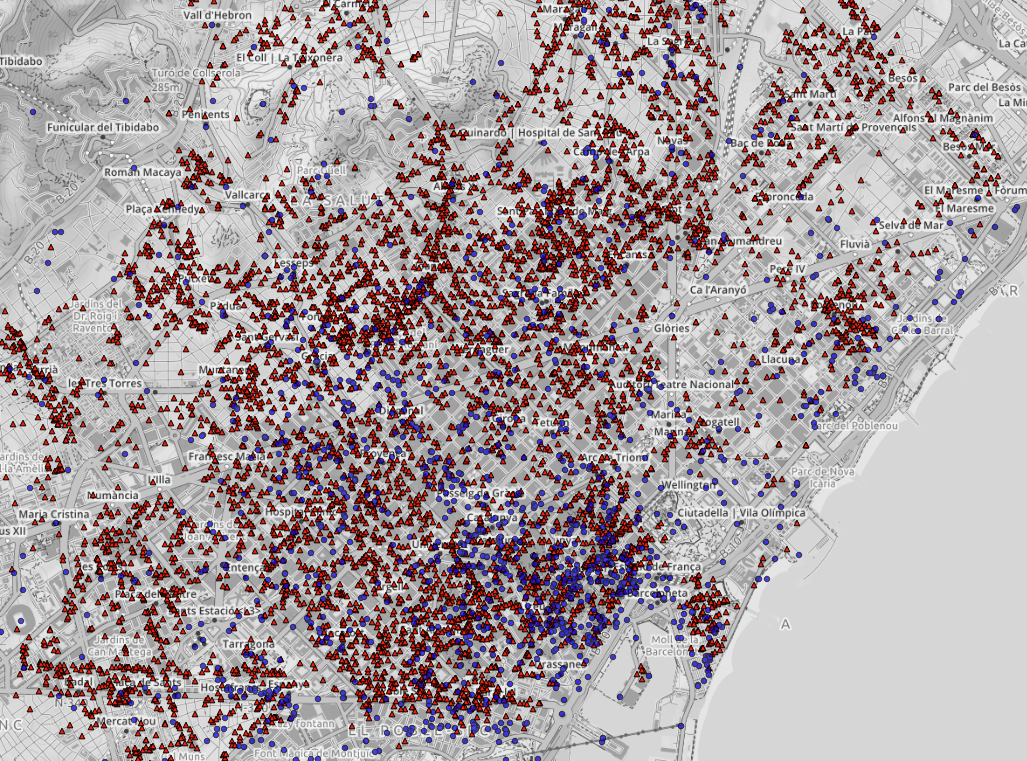}
  \caption{Figure 2. `Daily Purchases' (red) vs. `Attraction' (blue).}
  \label{fig:two}
\end{figure}

Figure~\ref{fig:two} depicts a visual example contrasting the places of the two different categories, Daily Purchases (presented with red dots) and Attractions (blue). It shows that there are more red triangles and that they are more dispersed, whereas the blue dots are more centered to the downtown area.

%% file: ubicomp_v1_nnm.tex
\subsection{All-Pair Mobility Modeling \& Model Auditing}

Associating neighborhood features with mobility is the key function of the framework that enables interpretations. However, understanding the complicated relations between the features and the mobility is a non-trivial task. Prior works on mobility models \cite{erlander1990gravity,simini2012universal} suggest that generalized rules that assume a certain relation (e.g., effect of distance on mobility) are limited in terms of  explaining the complexity of intra-urban mobility. Even if diverse features are available, the complexity of mobility is less likely to be explained in simple terms, for example, assuming higher mobility from an area with few schools to an area that have many schools. The features could interact in unexpected ways. In addition, mobility between a pair of areas could be influenced not only by the features of their own but also by the features of other areas surrounding the two. 

We approach to model the relations between the features and mobility using a multi-layer neural network. Instead of relying on pre-defined assumptions about the effect of the features, the method can perform tailored estimation of the features' importance as it learns from the actual mobility data of the target city. Furthermore, the approach can learn possible nonlinear interactions between the features by having multiple layers. 

Another strength of our approach is that it is designed to consider the effect of all other neighborhoods together when estimating the mobility flow between a pair of neighborhoods. Given a neighborhood, the model learns the mobility flow from the neighborhood to all others as a whole, instead of learning the flow to individual destinations independently. The prediction of the model for a neighborhood is a probability distribution, which represents the mobility flow between the neighborhood and all others.

We now describe the two steps of the approach, \textit{mobility modeling} and \textit{model auditing}, in detail. 

\paragraph{\textbf{Mobility Modeling}} 
Given a set of features $\textbf{x}_i$ for a neighborhood \textit{i}, we want a model to predict the flow probability distribution $\textbf{p}_i$, i.e., $\textbf{p}_i$ = Model($\textbf{x}_i$), where the \textit{j}-th component of $\textbf{p}_i$ represents the probability that a citizen from neighborhood \textit{i} moves to/from neighborhood \textit{j} in a given time frame. The direction of movement and the time frame are determined by the data set we develop (refer to the section Evaluation Design).

Since $\textbf{p}_i$ must contain real values, we could think of performing a regression for each of them. However, as $\textbf{p}_i$ represents a probability distribution, we want (i) to normalize the output to sum to 1 and (ii) to train the model considering the whole distribution $\textbf{p}_i$ and all the interactions between its components (as opposed to separately training one model for each component). The simplest model that fulfills these characteristics is the multivariate linear model with softmax output \cite{hastieelements}. This corresponds to the combination of a multivariate linear regression,

\begin{equation}
\textbf{y}_i = \textbf{W}\textbf{x}_i + \textbf{b}
\end{equation}

with a softmax function,
\begin{equation}
\textbf{p}_i = \frac{e^{\textbf{y}_i}}{\sum e^{\textbf{y}_i}}
\end{equation}
 
where \textbf{W} is a weight matrix, \textbf{b} corresponds to the bias vector, and the sum in the denominator of $\textbf{p}_i$ is taken over all the components of the vector.

More advanced models can be built upon since the simple model corresponds to a one-layer feed-forward neural network with a softmax activation \cite{goodfellow2016deep}. Thus, we can stack up several layers to obtain a (potentially more accurate) nonlinear model. We also explore this possibility by considering up to 4 layers with 100 rectified linear units \cite{glorot2010understanding} each. 

To train the models we use gradient descent and adapt the learning rate per dimension using ADADELTA \cite{zeiler2012adadelta}. We train for 3000 epochs using batches of 10 instances and shuffling. In order to avoid overfitting, we employ dropout \cite{srivastava2014dropout} with a probability of 0.5. In addition, we perform data augmentation \cite{goodfellow2016deep} by adding a 5\% Gaussian noise to the input \textbf{x}, which we previously normalize to have zero mean and unit standard deviation. Models' weights are initialized with the so-called Glorot initialization \cite{glorot2010understanding}.

\paragraph{\textbf{Model Auditing}}
For the estimation result of each neighborhood, the framework measures the importance of individual features through mean decrease accuracy (also known as permutation importance or direct influence) \cite{breiman2001random}. The general idea is to permute the values of each feature randomly, one at a time, and measure how much the permutation increases the error of the pre-trained model. Intuitively, the permutation of important variables should have a strong effect on model's accuracy, while permuting non-important variables should have little or no effect. For each feature, we measure the relative improvement (\%) of the estimation performance compared to when it is randomized.

%% file: ubicomp_v1_eval_design.tex
\section{Evaluation Design}
We focus our evaluation on the two primary functions of MobInsight: first, if the neighborhood features contribute to mobility estimation; second, if the feature analysis leads to sound interpretations specific to the target city. 

\subsection{Mobility Data of Barcelona}
We sample the mobility of Barcelona through a cell-phone network infrastructure. Cell phone networks are built using a set of base transceiver stations (BTS) that connect cell phones to the network. Each BTS has a latitude and a longitude, and gives coverage to an area called a cell. We follow the common practice \cite{frias2011agent, lenormand2015comparing} that assumes the cell of each BTS can be approximated by a two-dimensional non-overlapping polygon, and we use a Voronoi tessellation for the approximation. The location of the cell-phone user is assumed to be somewhere inside the cell. Note that no information about the exact position of users is known. 

The call data records (CDR) dataset used in this study contains all the phone calls, SMS, and MMS recorded by a major operator. The main fields of each CDR entry are: (1) a hashed ID of the originating cellphone number (2) that of the receiver (3) a time-stamp (when a call starts) (4) the duration of the call and (5) the BTS tower used.

The data was collected from the BTS towers located in Barcelona. The period of data collection was from Feb. 1 to Feb. 28, 2014. There were more than 700 active BTSs during the whole period, offering a sufficient level of segmentation of the city, much finer grained than the neighborhood-based division. In order to focus on real residents of the city and avoid tourist effects, we disregarded the records of roaming phones and those of pre-paid SIM cards. The data set had CDRs collected during one month, which account roughly for 2.5M unique phones and around 35M interactions. To preserve privacy, all the information is aggregated and encrypted. No contract or demographic data was considered, requested nor available for this study. Data collection and anonymization was done by a third party that was not involved in the analysis.

\subsection{Mobility Matrix of All-Pairs of Neighborhoods}
A mobility matrix, commonly called the \textit{O}-\textit{D} matrix \cite{frias2011agent}, characterizes the transitions of a population between different geographical regions representing the origin (\textit{O}) and destination (\textit{D}) of a route. Typically, \textit{O} and \textit{D} are the same set and represent the towns or neighborhoods of the geographical area under study. Each element of the matrix (\textit{i}, \textit{j}) defines the percentage or the total number of travels made to $\textit{D}_j$ by individuals who live in $\textit{O}_i$.

We construct the matrix at the neighborhood level from the CDR data. For this, we first apply a home detection algorithm that infers the users' home at a BTS level, and then group all the individuals whose home neighborhood is the same (group of BTSs within the region that defines the neighborhood). After that, the aggregated mobility of all the individuals of each neighborhood to other neighborhoods is estimated. The neighborhoods are defined by following the definition of the city municipality. To prevent noise, we only considered the users with at least 10 records in the data set. 

\paragraph{Home Detection} 
We used a simplified version of the algorithm presented in \cite{kung2014exploring}. For each user, we subsume the home as the neighborhood in which the most frequently used BTS cell is located, considering the records made between (1) Monday through Thursday from 20:00 to 08:00 and (2) Saturday and Sunday at any moment during the day.  If the second mostly used BTS has less than 80\% of the usage of the first one, we assume the first BTS to be the home location. Otherwise, we check the physical distance between the first and the second most used BTSs. If they are within 100 meters, we consider the first BTS as the home location. If the distance is larger, we assume that we cannot reliably identify any of them as the home and discard the users from the data. As a quick validation, we computed the Pearson correlation coefficient between our estimation of neighborhood population and that of the neighborhood census of Barcelona \cite{OpenBCN}, and observed 0.73 as the result.

\paragraph{Computation of Mobility Matrix} 
Once the home neighborhood of people is identified, the CDR entries from outside of it are considered as travelling samples. For each neighborhood, we count the residents' records found in all other neighborhoods. We only count the visits per day, so that if a user visits the same neighborhood several times during a day, it is counted as one visit. The final result is a 70x70 matrix that represents the frequency of visits to other neighborhoods during the considered time period. In the estimation experiment, the matrix is normalized per row to produce a normalized frequency distribution of travels to other neighborhoods.

\subsection{Mobility Estimation Setup}
We derive two estimation tasks from the mobility matrix: estimation of `To' and `From'. As the name indicates, the task `To' is the estimation of the relative frequency of travels made to other neighborhoods from the home neighborhood. On the other hand, `From' estimates the relative frequency of visits from all other neighborhoods to the home, which can be obtained by transposing the mobility matrix.

\subsubsection{Estimation Metric}
Due to the limited number of data points (\# of neighborhoods), we train and evaluate the model using leave-one-out cross-validation. Each iteration of the validation takes a particular neighborhood, which was left out in the training, and estimates the probability of travelling to all other neighborhood (or, for the `from' task, the probability of travels made to that neighborhood from all others). The quality of the estimation is measured using the Kullback-Leibler (KL) divergence \cite{kullback1951information} between the true (estimated) probability \textbf{q} and the predicted probability \textbf{p}.

\begin{equation}
D_{\text{KL}}(\textbf{p}_i|\textbf{q}_i) = \sum_i \textbf{p}_{ij} \log\left(\frac{\textbf{p}_{ij}}{\textbf{q}_{ij}}\right)
\end{equation}

The KL divergence measures the entropy increase caused by the estimated distribution relative to the ground-truth distribution.  In other words, it is the amount of information lost when \textbf{p} is used to approximate \textbf{q}. For a concise presentation, we report the average KL divergence across all the neighborhoods.

\subsubsection{Comparison Methods}
We briefly describe the comparison methods below.

\begin{itemize}
\item{\textbf{Random Model}}:
This model produces a random probability distribution at all iterations of the cross-validation. It gives us an indication of the chance level in our tasks.

\item{\textbf{Average-based model}}:
At each iteration, this model takes the average mobility of the other neighborhoods. More specifically, to estimate the mobility matrix \textbf{M}, the row $\textbf{m}_i$ of the neighborhood \textit{i} is estimated at each iteration. Each element of the row is calculated as $\textbf{m}_{ij}=\frac{1}{\textbf{N}} \sum_{k \in N} \textbf{m}_{kj}$, where \textbf{N} is the set of the neighborhoods in the training set. 

\item{\textbf{Gravity model} \cite{erlander1990gravity}}: 
This model uses the population of neighborhoods, \textit{H}, and the distance information between them, \textit{d}. 

\begin{equation}
\textbf{M}_{i,j} = g \frac {\textit{H}_i \textit{H}_j} {d^2_{i,j}}
\end{equation}

We acquired the population data from the open government data of Barcelona \cite{OpenBCN} and approximated the distance with the transportation time between all pairs of neighborhoods returned from Google Maps. The scaling constant \textit{g} is optimized to the value which produces the minimum KL-divergence (a grid search over the full data set is performed, hence it is an optimistic performance estimate \cite{hastieelements}). 

\item{\textbf{Proposed Model} (`NF\_Dist')}:
It uses neighborhood features (NF) and distance information (Dist). 

\item{\textbf{Proposed Model without open government data} (`NF\_woPub\_Dist')}:
In order to understand the importance of having a less-skewed data set covering the places of the public sector, we also create a comparison method employing the same methodology, except that we exclude the open government data set. 

\item{\textbf{Pairwise Model}}:
Through this model, we intend to see the performance when individual mobility flows are learned and predicted independently, not as a whole. While all configurations are kept as the same as in the proposed model, a separate model is trained for each element $\textbf{p}_{ij}$ of the probability distribution vector $\textbf{p}_i$ explained in section 3.2. 

\item{\textbf{Vector Distance Model} (`VecDist')}:
This method uses a simple vector-based distance measure for the prediction, which does not learn the effect of features from data. The prediction is based on the neighborhood features (NF) and the distance (d), where cosine similarity is used for measuring the difference of the neighborhood features.

\begin{equation}
\textbf{M}_{i,j}= \frac {(1-cossim(\textbf{NF}_i, \textbf{NF}_j))} {\textit{d}_{ij}}
\end{equation}

\end{itemize}

\subsection{Approach to Evaluation of Mobility Interpretations}
In contrast to the mobility estimation performance, the interpretations are inherently qualitative and subtle. The correctness of an interpretation is not always straightforward and it is hard to clearly define the scope of right answers. In addition, the interpretations of our interest deal with the mobility patterns at a city-scale, making it difficult to conduct surveys or find experts who are familiar with such large-scale patterns. 

While admitting the difficulties of conducting a thorough evaluation, we focus on delineating the types of interpretations that are enabled by MobInsight, and discuss why they are difficult to make with conventional approaches. While we use empirical examples to elaborate on each type of interpretation, instead of making arbitrary choices of the examples, the choices are made based on the estimation performance and the clarity of feature analysis. For example, we look into the neighborhoods where the estimation performance of the framework is much higher than the baseline, and those that have distinguished features contributing to the estimation performance. 

We also discuss the validity of the individual examples using various resources about the urbanism of Barcelona including the census, socio-economic data, urban development projects, related articles in the encyclopedia of Barcelona, etc. 

%% file: ubicomp_v1_results_est.tex
\section{Results and Discussion}
\subsection{Mobility Estimation Performance}
Table~\ref{tbl:three} provides an overview of the performance of different models for the two tasks, \textit{i.e.}, To/From. The proposed model using only one layer (NF\_Dist) achieves around 35\% relative improvement with respect to the Average model, and 30\% with respect to the Gravity model. To elaborate further on the improvement in an intuitive way, we use an example estimation for a neighborhood that showed a KL-divergence improvement of 0.08 (absolute) over the Gravity model. For this neighborhood, we had 150k travels to other neighborhoods and our model reduced the total estimation error from 66k to 42k travels. 

\begin{table}
  \includegraphics[scale=0.45]{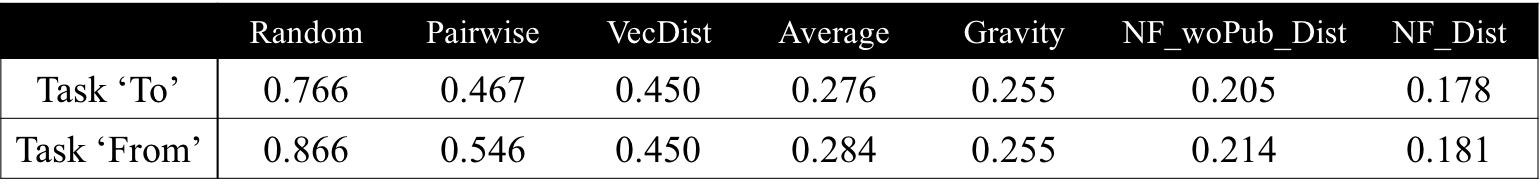}
  \caption{Performance comparison.}
  \label{tbl:three}
\end{table}

The improvement is promising, especially considering that we used a lower resolution for the distance information than the one used for the Gravity model. The full distance information between all neighborhood pairs produces too many additional features (70) with respect to the number of neighborhood features (18) and the number of available training data. Thus, we lowered the number of features by only using the (latitude, longitude) coordinates of the neighborhood center points. The coordinates approximate the distance since the model is able to compute the difference of longitude and latitude respectively and combine them.

Apart from the main implication that the neighborhood features contribute to explaining the mobility, this result offers multiple implications for mobility estimation applications. First, the use of neighborhood features instead of population can greatly reduce the time required for an accurate estimation. While urban spaces evolve over time, the consequent changes of population can happen slowly. On the other hand, the changes in the neighborhoods are likely to be updated to the web much before the population adapts. This opens the opportunity to obtain dynamic estimations in a timely manner. 

Second, it enables the prediction of mobility while making hypothetical assumptions. The neighborhood features support hypothesizing changes of places in different areas and obtaining a corresponding estimation of mobility. Developing hypotheses in terms of places would be easier and more realistic than those assuming a change of population.
 
Third, the comparison between NF\_Dist and NF\_woPub\_Dist supports our intuition that it is important to include the places that are neglected in commercial sources in understanding mobility. The proposed model shows around 15\% of improvement over NF\_woPub\_Dist. As many recent works in urban informatics consider social media as a main data source, our result offers a useful implication to such work about the possible limitation of the source, especially if the sources are used in the context of understanding mobility. 

The comparison of the proposed model to VecDist shows the importance of tailoring the model based on the actual data, and the comparison to Pairwise reveals the importance of considering the effect of all neighborhoods together. We further elaborate on these points  through example interpretations in the next section. 

\begin{table}
  \includegraphics[scale=0.45]{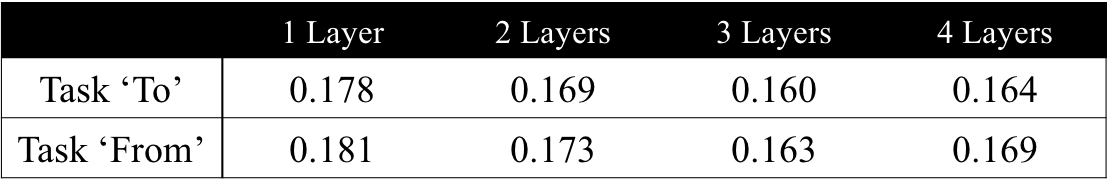}
  \caption{Effect of the number of layers.}
  \label{tbl:four}
\end{table}

In addition to the above comparisons, we observed additional improvement of our model when a few more layers were added to the neural network (Table ~\ref{tbl:four}; notice that the 1-layer model corresponds to NF\_Dist in Table ~\ref{tbl:three}). The improvement was observed until the third layer was added and started degrading when the fourth layer was added, possibly because of the limited training data.

%% file: ubicomp_v1_results_int.tex
\subsection{Interpretations of Barcelona's Mobility}
\subsubsection{Larger Space of Interpretation}
An obvious benefit of the framework is the availability of diverse features that offer a plausible interpretation for the cases that simpler models could not explain. We elaborate on this point using a visualization that illustrates an example. 

\begin{figure}
  \includegraphics[scale=0.5]{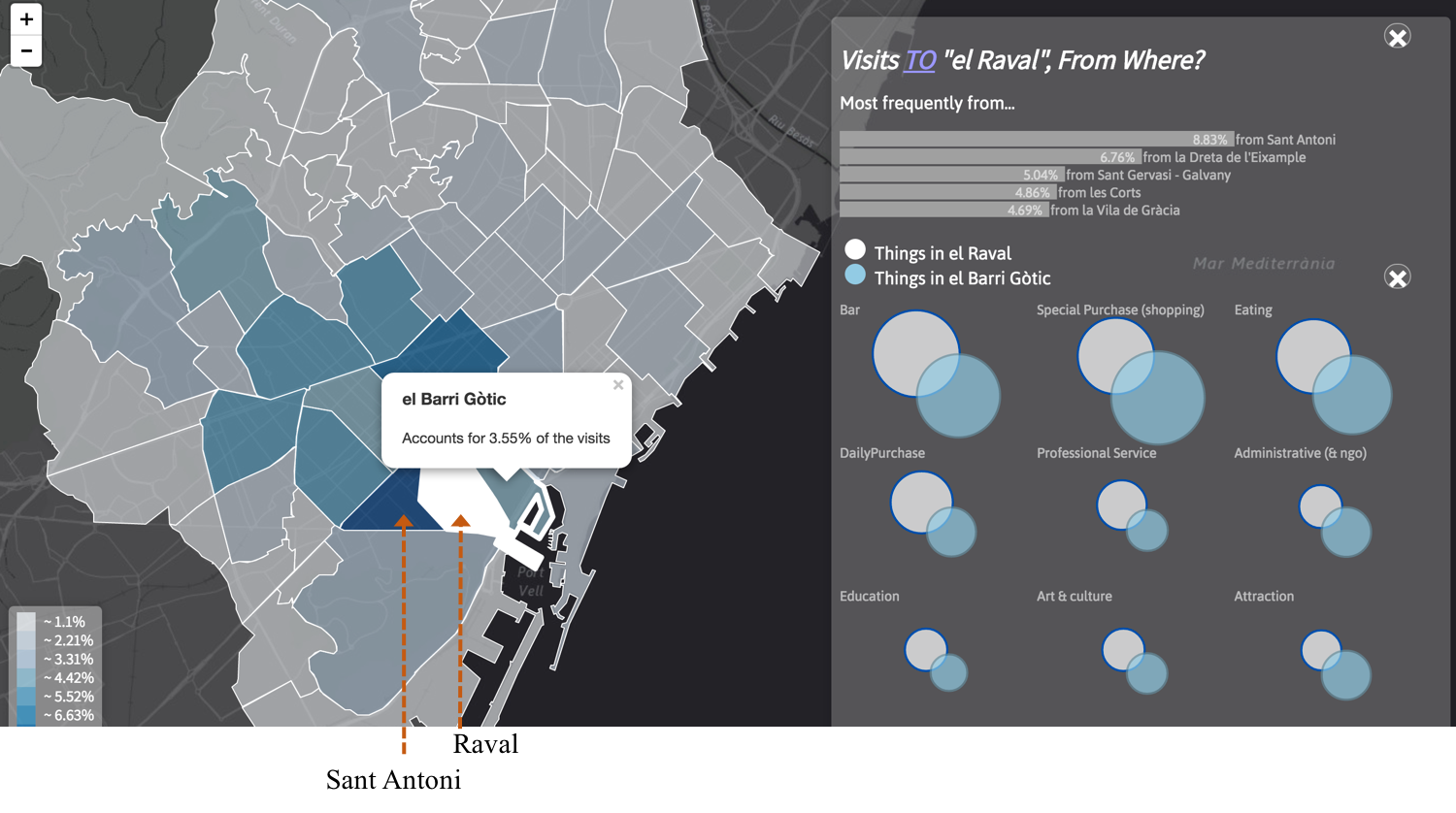}
  \caption{Example interpretation through visual exploration of the mobility and neighborhood features.}
  \label{fig:three}
\end{figure}

Figure ~\ref{fig:three} depicts the mobility towards a selected neighborhood (Raval, highlighted in white) over a map, and a bubble chart that shows the neighborhood features of Raval and Barri G\'otic, which is selected for feature comparison. The color-coding of the map is based on the frequency of mobility towards Raval, where darker colors represent more frequent visits. 

An interesting point is the significant difference between the amount of visits made from Sant Antoni and that from Barri G\'otic. However, a simple mobility model that does not consider local features would not explain the drastic difference given the close distances between the two neighborhoods and Raval. 

The diverse set of features help make sense of the different relationship between the neighborhoods and obtain a plausible explanation. In accordance with the fact that Raval is one of the city center area where people often visit for nightlife and shopping, the white bubbles that represent the features of Raval have a greater size for the features Bar, Special purchase (shopping), and Eating. 

On the other hand, the blue bubbles that represent the features of G\'otic also reveals that G\'otic is another neighborhood composing the city center area which has many places for nightlife and attractions. As for the three main features of Raval mentioned above, the corresponding bubbles of the two neighborhoods show comparable sizes. It enables the interpretation that the people of G\'otic might find Raval less attractive despite the close distance, and other factors than nightlife or attraction would be important for them to visit other neighborhoods.

\subsubsection{Interpretations Tailored to Target Areas}
As mentioned, the framework identifies the important features for explaining the mobility of each neighborhood. We frequently observed that the identified features are different from those that would have been identified by other simple approaches, such as generalized models emphasizing the role of work and home places, or vector similarity measures (\textit{e.g.}, cosine similarity). In order to save space, we use a number of examples instead of reporting the feature analysis of all the neighborhoods.

The result for the neighborhood Pedralbes provides a typical example. Pedralbes is a well known wealthy uptown residential area, described as a neighborhood that `stands out from others in terms of socio-economic class' \cite{wiki:Urbanismo}. It is also described to have `the service sector as its main economic activity and hosts financial institutions and office centers'.  Indeed, the feature analysis showed high importance for `Professional services', and `Offices', implying that they are critical for estimating the mobility of Pedralbes. In addition, the feature `Education' also showed high importance as Pedralbes hosts many private and international schools, and university campuses.

While the identified features match with the urbanism of Pedralbes, another important aspect of the result is that the framework did not put high importance to some features that would have been identified to be important by vector similarity measures. Although the features `administrative offices', `special purchases', and `leisure' were distant from the average of other neighborhoods, MobInsight assessed the effect of these features to be negligible.  

As for another example, in Barceloneta, the importance values were exceptionally skewed to `Club' and `Eating'. This seemed trivial as the area is described to be ``famous for its beach, restaurants, and nightclubs along the boardwalk'' \cite{wiki:Barceloneta}. However, we found this example interesting since there are many areas with diverse nightlife and restaurants options (\textit{e.g.}, other neighborhoods in the city center), and the exceptional skew of Barceloneta indicates, on the other hand, that the other features do not contribute much. A plausible explanation for the skew could be that the area went through a profound transformation during the urban project near the 1992 Olympics which aimed to strengthen its recreational function. The area is known to be struggling over gentrification after the project \cite{wiki:La_Barceloneta}. Although there were a number of features whose value were significantly distant from the average of other neighborhoods, the framework put much less importance to them compared to the two key features. 

Building upon this point, we observe if the skew of feature importance is an indicator of a certain residential quality. Jane Jacobs, in her famous book ``The life and death of great American cities'' \cite{jacobs1961death}, argues about essential conditions for a lively neighborhood, and emphasizes the importance of having a mix of diverse functions in a neighborhood. Inspired by her argument and the recent evaluation of the argument conducted by De Nadai et al. with a few Italian cities \cite{de2016death}, we conducted a correlation analysis between the degree of skew and population. We first measured the variance of feature importance for each neighborhood, assuming that the neighborhoods with diverse features of high weight will show low variance. Then, we analyzed the correlation between the computed variance and the neighborhood's population. Interestingly, we observed a significant inverse correlation between the variables (Pearson coefficient=-0.3, $\textit{p}<0.05$). Such a result could be capturing the preference for neighborhoods with diverse functions, which supports Jane Jacobs' original argument. 

\subsubsection{Insights beyond Individual Neighborhood Pairs}
As mentioned, the framework views the mobility flow from one neighborhood to all others as a whole rather than to look at individual flows separately. While this leads to superior performance overall than the baselines that models the mobility flows independently (\textit{e.g.}, gravity model), we elaborate on two simple examples that demonstrates the importance of this point. 

\begin{figure}
  \includegraphics[scale=0.5]{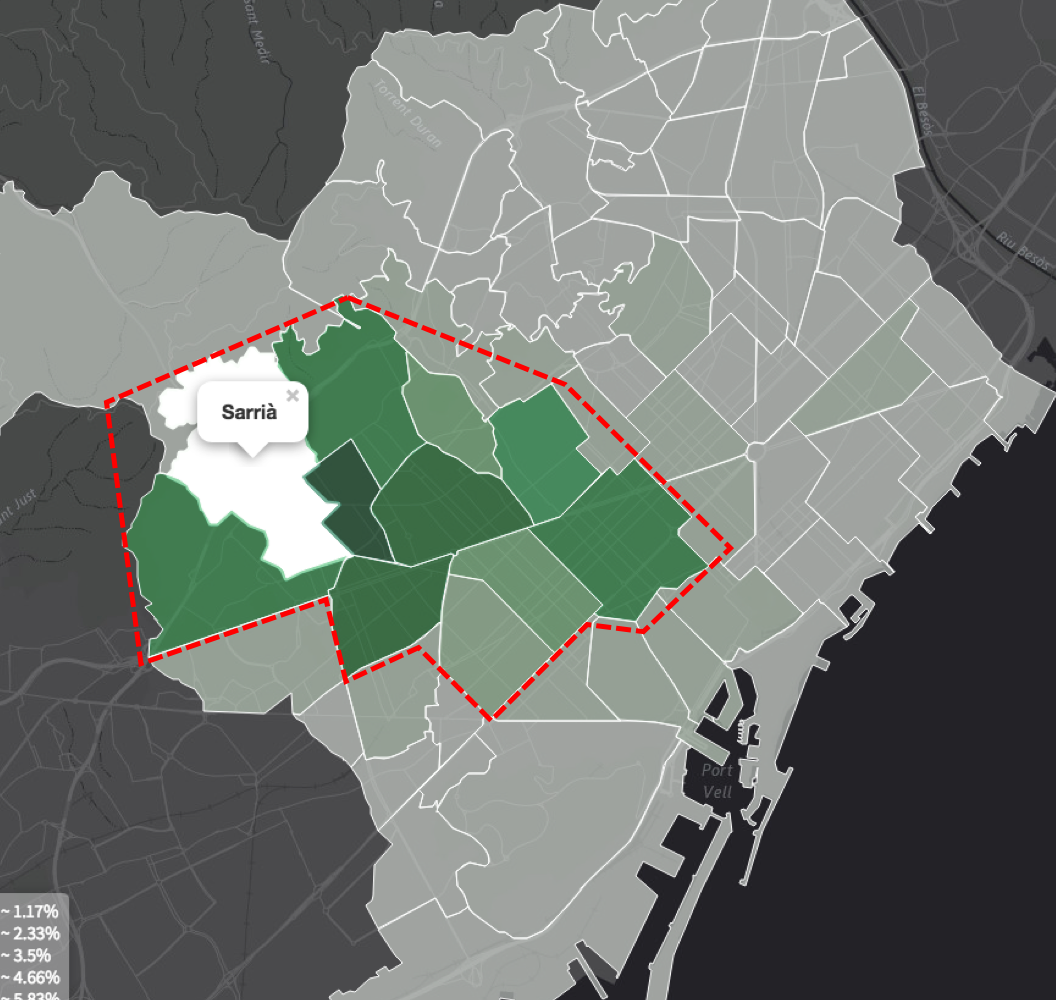}
  \caption{Mobility from Sarri\'a to other Neighborhoods.}
  \label{fig:four}
\end{figure}

Figure ~\ref{fig:four} shows the ground truth mobility flow from one neighborhood, Sarri\'a, to all other neighborhoods. It shows that the mobility is strongly clustered to a number of neighborhoods within the marked area. Each of the 10  neighborhoods in the marked area absorbed 6\% of the mobility from Sarri\'a on average, whereas all the others absorbed less than 2\%. This contrast implies that, for example, the infrequent mobility from Sarri\'a to a  neighborhood outside the marked area cannot be understood by only looking into the relationship between the two. Rather, it is important to take into account the strong ties that Sarri\'a has with the neighborhoods in the marked area.  

In addition to achieving better estimation of such contrasting mobility flows, MobInsight's feature analysis suggests a possible interpretation. For Sarri\'a and the surrounding neighborhoods, MobInsight frequently assigned high importance to the features `Education', `Health', `Offices', and `Leisure', whereas those features had much less importance for the neighborhoods outside the marked border. The contrast leads to the speculation about the possible effect of socio-economic differences between the two parts of the city on the mobility. According to Barcelona's official statistics \cite{BCN_STATS}, the neighborhood cluster around Sarri\'a is the richest in the city with family incomes between 1.8 and 2.5 times higher than that of the neighborhoods outside. The clustered mobility could be implying socio-economic homophily between the affluent neighborhoods, reinforced by the fact that many of the education and health facilities in those neighborhoods are private. It also extends the findings made about the relation between mobility and social deprivation in the prior works \cite{lathia2012hidden,smith2013finger}.

\begin{figure}
  \includegraphics[scale=0.5]{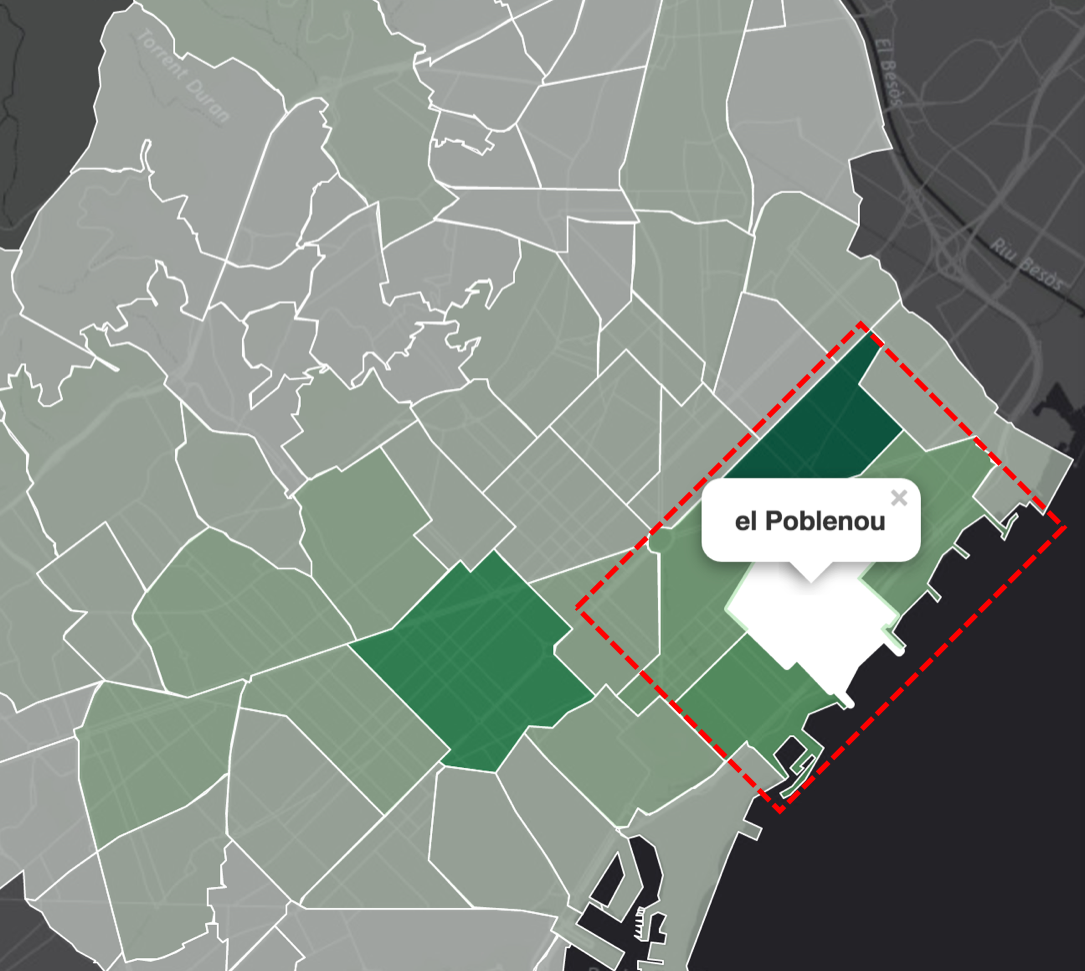}
  \caption{Mobility from Poblenou to other Neighborhoods.}
  \label{fig:five}
\end{figure}

A similar contrast is found between the neighborhoods around Poblenou and the rest, especially the ones on the north (Figure ~\ref{fig:five}). The feature analysis of these neighborhoods around Poblenou commonly showed an exceptionally high importance for `Office'. We believe the result is capturing the specialized characteristic of the area resulting from an urban project conducted in the early 2000s. The area was suffering from deindustrialization and the project transformed the area from an industry zone of factories to a highly specialized zone for emerging industries, which now hosts numerous technology companies (over 7000 businesses), and foreign employees \cite{fugueras2005enciclopèdia}.

%% file: ubicomp_v1_conclusion_refs.tex
\section{Conclusions and Future Work}
In this paper, we have presented MobInsight, a framework that supports deeper interpretations of urban mobility specific to the target city. It takes advantage of the interpretable features produced by holistic semantic aggregation, the method we create for neighborhood profiling. The method thoroughly identifies existing places in the neighborhoods and extracts the semantic meanings from the annotations left on them. The framework comprehensively analyzes how the features affect the mobility by building mobility models and performing model auditing. We evaluate the framework with the mobility data of Barcelona and elaborate on three types of interpretations that touch the urbanism of Barcelona. 

Our on-going works include creating and testing new semantic features for neighborhood profiling. There are additional data that are already collected but not used currently, e.g., scores given to places, full-text reviews, etc. We believe further improvement can be made by developing new features from them, such as the popularity/quality of places, different preferences between age groups or cultural background, etc. Another future direction is to expand the analysis to other cities and make more site-specific interpretations through comparison. 

As mentioned, developing a methodology for a thorough evaluation of the interpretations is a challenging future work. We believe it is a large topic which should be covered in a separate study as it requires thoughtful design in terms of choosing developing ground-truth and evaluation tasks, comparison methods, and finding qualified experts as evaluators. We will explore various ideas in qualitative methodologies and also the recent advances in crowd-sourced evaluation methods. 

% Bibliography
\bibliographystyle{ACM-Reference-Format}
\bibliography{ubicomp_v1_references}